\documentclass[11pt]{article}

\usepackage{amsmath}

\begin{document}

\title{Lorentz group theory and polarization of the light \\
V.M. Red'kov\footnote{redkov@dragon.bas-net.by}
  \\[3mm]
{\small B.I. Stepanov Institute of Physics\\  National Academy of
Sciences of Belarus}
}

\maketitle

\begin{quotation}

Some facts of the theory of the Lorentz group
are specified for looking at the problems of light
polarization optics in the frames of vector Stokes-Mueller  and
spinor Jones  formalism.
In view of great differences between  properties of isotropic
and time-like vectors
 in Special
Relativity we should expect  principal differences in
describing completely polarized and partly polarized light.
In particular, substantial differences are
revealed when turning to   spinor  techniques in the context of the polarized light.

 Because   Jones complex
formalism has close relation to spinor objects of the Lorentz
group, within  the field of the light polarization  we  could have
physical realizations on the optical desk of some subtle
topological distinctions between  orthogonal $L_{+}^{\uparrow} =
SO_{0}(3.1)$ and spinor $SL(2.C)$ groups. These topological
differences  of the groups find their corollaries in the  problem
of the  so-called spinor structure of physical space-time, some new points are considered.

\end{quotation}


\section{Introduction}

The main line of evolution in
theoretical methods of polarization
 optics seems to be quite independent of that
in relativistic symmetry methods,  developed, for example, in
particle physics. By many authors it was  noticed that  these two
branches of physics employ, in fact, the same mathematical
technique\footnote{The bibliography on the subject is large
enough -- see  the big list of references in  \cite{2009-Red'kov}.}, only with occasionally motivated   distinctions  in notation and
physical accents. So, in this part,  the present article is one
other appeal: instead of remaking the same things in  different
embodiment, it is better to work out  and adopt a unique
mathematical language. In so doing we might see the more unity and
simplicity in symmetry aspects of particle physics and optics of
the polarized light. Because   Jones complex formalism has close
relation to spinor objects of the Lorentz group, within  the field
of the light polarization  we  could have physical realizations on
the optical desk of some subtle  topological distinctions between
orthogonal $L_{+}^{\uparrow} = SO_{0}(3.1)$ and spinor $SL(2.C)$
groups. These topological differences  of the groups find their
corollaries in the  problem of the  so-called spinor structure of
physical space-time.

In the paper a     technique of working with the
Lorentz is used, the  systematic
 construction  of of that was  given by Fedorov \cite{Fedorov-1980},
  which  closely related to
 studies of Einstein and Mayer
 \cite{Einstein-Mayer(1), Einstein-Mayer(2), Einstein-Mayer(3)} on semi-vectors,
and  also closely  related to quaternionic approach
\cite{Berezin-Kurochkin-Tolkachev-1989, 2005-Gsponer-Hurni(1),
 2005-Gsponer-Hurni(2)}.
This technique  is specified for looking at the problems of light
polarization optics in the frames of vector Stokes-Mueller  and
spinor Jones  formalism.

Remembering on great differences between  properties of isotropic
and time-like vectors
 in Special
Relativity we should expect the same principal differences in
describing polarized and partly polarized light. So below we will
be considering  these two cases  separately:  a polarized light
and  a partly polarized light. In particular, substantial differences will be
revealed when turning to   spinor techniques.

In the first place, one may restrict oneself to rotation subgroup of the Lorentz group and
study   corresponding manifestation optical devices in such a
non-relativistic limit. So, the main idea  in the  optical context
should be to derive as much as possible from conventional theory
of
 the rotation group
$SO(3.R)$ and its spinor covering, unitary group $SU(2)$; turning
to spinors of  $SU(2)$ latter immediately leads us to an
additional insight to complex Jones formalism in polarization
optics. Then we consider  application of relativistic group $SL(2.C)$ when
naturally  bi-spinor of Jones type and 4-vector and 4-tensor of Stokes  type arise.

Also we will describe some possibilities to apply in optical context the well-known ideas
  closely
related to it ideas on spinor space-time structure \cite{1984-Penrose}.

This  paper only partly represents a  more comprehensive treatment given in
\cite{2009-Red'kov}.

\section{Polarization of the light,  Mueller  4-vector  formalism}

To elucidate how mathematical facts on rotation and Lorentz group
may be applied to problems in polarization optics,
and also what problems of this field wait to be solved, let  us
start with some basic definitions  concerning the polarization of
the light (at this we have used \cite{1992-Snopko}, though it
might be another from many).

For a plane  electromagnetic wave spreading along the axis $z$, in
an arbitrary fixed point $z$ we have ($M,N$ are amplitudes of two
electric components, $\Delta$  is a phase shift)
\begin{eqnarray}
E^{1} = N \cos \omega t \; , \qquad E^{2} = M \cos (\omega t +
\Delta ) \; , \qquad E^{3} = 0\; , \nonumber
\\
N \geq 0 \; ,  \qquad M \geq 0\; , \qquad  \Delta \in [- \pi , +
\pi ] \; ; \label{3.1}
\end{eqnarray}

\noindent four Stokes parameters $(S_{a}) =(I, S^{1},
S^{2},S^{3})$ are determined  by
\begin{eqnarray}
I =\; < E_{1}^{2} + E_{2}^{2} > \; , \qquad S^{3} = \;< E_{1}^{2}
- E_{2}^{2} >  \; , \nonumber
\\
S^{1} = \;< 2E_{1}E_{2} \; \cos \Delta > \; , \qquad  S^{1} = \;<
2E_{1}E_{2}\;  \sin \Delta > \; ; \label{3.2}
\end{eqnarray}

\noindent the symbol $ < ...>$  stands for  averaging in time.

If
the amplitudes $N,M$ and  the phase shift $\Delta$ do not depend
on time in measuring process, the Stokes parameters equal to
\begin{eqnarray}
S^{0} _{pol} = I_{pol} = N^{2} + M^{2} \;  , \qquad
S^{3}_{pol} = N^{2} - M^{2} \; , \nonumber
\\
S^{1}_{pol} = 2NM\;\cos \Delta  \; , \qquad S^{2} _{pol}=
2NM\; \sin \Delta  \; , \label{3.3}
\end{eqnarray}

\noindent and the identity holds
\begin{eqnarray}
S_{a} S^{a} = (S^{0}_{pol})^{2} - S^{j}_{pol} S^{j}_{pol} =
I^{2}_{pol} - {\bf S}^{2}_{pol} = 0 \;  ; \label{3.4}
\end{eqnarray}

\noindent that is ${\bf S} = I_{pol} \; {\bf n}$. In other
words, for a  completely polarized light, the Stoke 4-vector is
isotropic.
 For a natural
(non-polarized)  light, Stokes parameters are trivial:
\begin{eqnarray}
S^{a}_{non-pol}  = ( I_{non-pol} , 0, 0, 0)\;  . \nonumber
\end{eqnarray}

\noindent
When summing  two non-coherent light waves,  their Stokes
parameters behave in accordance with the linear law: $ I_{(1)} +
I_{(2)} \; , \;  {\bf S} _{(1)} +  {\bf S}_{(2)} $. In particular,
a partly polarized light can be obtained as a linear  sum of
natural and completely polarized light:
\begin{eqnarray}
S^{a} _{non-pol} = (I_{non-pol} , 0, 0, 0)\;, \qquad
S^{a}_{pol} = ( I_{pol},  I_{pol} \; {\bf n} )\;, \nonumber
\\[3mm]
S^{a} =  (\; I_{non-pol}  +
I_{pol} \; )\;  \left (1 , { I_{pol}  \over  I_{non-pol}  +
I_{pol} } \; {\bf n}  \right ) \; . \nonumber
\end{eqnarray}

\noindent With notation
\begin{eqnarray}
I =  I_{non-pol}  +  I_{pol}  \; , \qquad p = { I_{pol}
\over  I_{non-pol}  +  I_{pol}  }\;  , \label{3.6}
\end{eqnarray}

\noindent for the Stokes vector of the partly polarized light we
have
\begin{eqnarray}
S^{a} =  ( I , \; I\; p \; {\bf n} ) \; , \qquad   S_{a}S^{a} =
I^{2}(1 - p^{2}) \geq 0 \; , \label{3.7}
\end{eqnarray}

\noindent where  $I$ is a general  intensity, $p$ is a degree of
polarization which runs within  $[0,\; 1]$ interval: $0 \leq p
\leq 1 $,  ${\bf n}$ stands for any 3-vector.

Behavior of Stokes 4-vectors  for polarized and partly polarized
light under acting optics devices
 may  be considered
  as isomorphic to behavior of  respectively isotropic and time-like vectors  with respect to Lorentz group
  in Special Relativity. This simple observation leads to many consequences, some of them will be discussed below.

\section{Polarized light  and Jones  formalism,
restriction to
   symmetry $SU(2)$, and two  sorts of  non-relativistic Stokes 3-vectors
}

Now let us consider the Jones formalism and its connection with
spinors for rotation and Lorentz groups. It is convenient to start
with a  relativistic 2-spinor $\psi$, representation of the
special linear  group $GL(2.C)$, covering for the Lorentz group $L_{+}^{\uparrow}$:
\begin{eqnarray}
\Psi = \left | \begin{array}{c} \psi^{1} \\ \psi ^{2}
\end{array} \right | \; , \qquad  \Psi ' = B(k) \Psi \; , \qquad  B(k) \in
SL(2.C)\; , \nonumber
\\
  B(k) = k_{0} + k_{j} \sigma^{j} \;  , \qquad
 \mbox{det} = k_{0}^{2}- {\bf k}^{2}  = 1 \; .
\label{4.12}
\end{eqnarray}

From the spinor $\psi$ one may construct a 2-rank spinor $\Psi
\otimes \psi^{*} $,   which in turn can be resolved in terms of
Pauli matrices (we  need two sets: $\sigma^{a} = (I,
\sigma^{j})$ and   $\bar{\sigma}^{a} = (I, -\sigma^{j})$):
\begin{eqnarray}
\Psi \otimes \Psi^{*} = {1 \over 2} \;( S_{a} \; \bar{\sigma}^{a}
)  = {1 \over 2} \; (S_{0} - S_{j}\; \sigma^{j} )\; . \label{4.13}
\end{eqnarray}

\noindent
The spinor nature of $\Psi$  generates a definite (Lorentz)
transformation law for $S_{a}$:
\begin{eqnarray}
 S\;'_{a} \; \bar{\sigma}^{a}
= S_{a} \;  B(k)   \bar{\sigma}^{a} B^{+}(k) \; , \label{4.14}
\end{eqnarray}

\noindent
or with the use of the  well-known relation in the theory of the
Lorentz group:
\begin{eqnarray}
B(k)   \bar{\sigma}^{a} B^{+}(k) =  \bar{\sigma}^{b}
L_{b}^{\;\;\;\;a}  \qquad \Longrightarrow \qquad S'_{b} =
L_{b}^{\;\;\;a} \; S_{a} \nonumber
\\
L^{\;\; a}_{b} (k,\; k^{*}) = \bar{\delta }^{c}_{b} \; \left [ \;
- \delta^{a}_{c} \; k^{n} \; k^{*}_{n} \;  + \;  k_{c} \; k^{a*}
\; + \; k^{*}_{c} \; k^{a}\; + \; i\; \epsilon ^{\;\;anm}_{c}\;
k_{n} \; k^{*}_{m} \; \right ] \;  , \label{4.15}
\\
\mbox{modified Kronecker symbol} \qquad
\bar{\delta }^{c}_{b} =
\left \{
\begin{array}{l}
+1 , \; \;  c = b = 0 \; ; \\
-1 , \; \;  c = b = 1, \; 2,\; 3 \; .
\end{array} \right.
\nonumber
\end{eqnarray}

\noindent Thus,  spinor transformation $B(k)$ for spinor $\psi$
generates  vector transformation  $L_{b}^{\;\;\;a} (k, k^{*})$.
Different in sign spinor matrices, $\pm B$, lead  to one the same
matrix $L$. If we restrict ourselves to the case of $SU(2)$ group,
we get
\begin{eqnarray}
k_{0}= n_{0} \; , \qquad k_{j} = - i n_{j} \; , \qquad n_{0}^{2}
+ n_{j}n_{j} = +1 \; , \qquad B (n) = n_{0} -i n_{j} \sigma_{j} \;
, \nonumber
\\
L(\pm n) =
 \left | \begin{array}{rrrr}
1  & 0  &  0  & 0  \\
0 & 1  -2 (n_{2}^{2}+n_{3}^{2})  & -2n_{0}n_{3}+2n_{1}n_{2}  & 2n_{0}n_{2}+ 2n_{1}n_{3}  \\
0 & 2n_{0}n_{3}+2n_{1}n_{2}  &  1  - 2( n_{1}^{2} +n_{3}^{2} )  & - 2n_{0}n_{1} +2n_{2}n_{3}  \\
0 & -2n_{0}n_{2}+2n_{1}n_{3}  &  2n_{0}n_{1}+2n_{2}n_{3}  &  1 -2(
n_{1}^{2}+n_{2}^{2})
\end{array}  \right | \; .
\label{su-2'}
\end{eqnarray}

Let us introduce a polarization  Jones spinor $\Psi$:
\begin{eqnarray}
\Psi =
\left | \begin{array}{r} N e^{i\alpha}  \\ M   e^{i\beta}
\end{array} \right | , \qquad   \psi \otimes \psi^{*} \qquad
\qquad
\nonumber \\
= \left | \begin{array}{cc}
N^{2} & NM e^{-i(\beta-\alpha)} \\
NM e^{+i(\beta-\alpha)} & M^{2}
\end{array} \right |
=
 {1 \over 2} \left | \begin{array}{cc}
S^{0} + S^{3} & S^{1} - i S^{2}     \\
 S^{1} + i S^{2}           &  S^{0}-S^{3}
 \end{array} \right |  ,
\label{4.19}
\end{eqnarray}

\noindent that is
\begin{eqnarray}
  S^{1} = 2NM \cos (\beta - \alpha)\; , \qquad S^{2} = 2NM \sin (\beta - \alpha)\; ,
\nonumber
 \\
 S^{3} = N^{2} - M^{2} \; , \qquad S^{0} = N^{2} + M^{2}  = + \sqrt { S_{1}^{2} +S_{2}^{2} + S_{3}^{2} }\;   .
\label{4.21}
\end{eqnarray}

\noindent Formulas (\ref{4.21})  should be compared with eqs. (\ref{3.3})
\begin{eqnarray}
S^{0} = N^{2} + M^{2}  = + \sqrt { S_{1}^{2} +S_{2}^{2} +
S_{3}^{2} }\;  ,  \qquad  S^{3} = N^{2} - M^{2} \;.
 \nonumber
 \\
 S^{1} = 2NM\;\cos \Delta  \; , \qquad  S^{2} = 2NM\; \sin \Delta
\;  . \label{4.22}
\end{eqnarray}

\noindent They coincide if $(\beta - \alpha)= \Delta$. Instead of
$\alpha , \beta$ it is convenient introduce new variables:
\begin{eqnarray}
\Delta = \beta - \alpha \;, \qquad  \gamma =  \beta + \alpha \; ,
\nonumber
\label{4.23}
\end{eqnarray}

\noindent correspondingly the spinor $\Psi$  looks
(\ref{4.12})
\begin{eqnarray}
\Psi = e^{i\gamma /2} \; \left | \begin{array}{r} N \;e^{- i\Delta
/2 }
 \\ M\;  e^{+i\Delta/2}
\end{array} \right |  =
e^{i\gamma /2} \; \left | \begin{array}{r} \sqrt{(S+S^{3} )/2}\;
\;e^{- i\Delta /2 }
 \\ \sqrt{(S-S^{3} )/2}\; \;  e^{+i\Delta/2}
\end{array} \right |
. \label{4.24}
\end{eqnarray}

Let us write down the  inverse relations to
(\ref{4.22}),  they are
\begin{eqnarray}
2N^{2} =  S+S^{3}  \;  , \qquad  2 M^{2}  =  S - S^{3}  \; ,
\qquad \Delta = \mbox{arctg}\; {S^{2} \over S^{1}} \; ;
\label{4.26}
\end{eqnarray}

\noindent  these correlate with  the known  formulas  defining
parabolic  coordinates in Euclidean  space:
\begin{eqnarray}
\xi = r + z \; , \qquad \eta = r - z \; , \qquad \phi =
\mbox{arctg} \; {y \over x} \; . \label{4.27}
\end{eqnarray}

\noindent
  Evidently, we have isomorphism between parameters of the Jones spinor
$(N,M, \Delta )$  and parabolic coordinate $(\xi, \eta, \delta)$
in effective space  of Stokes 3-vector $(S^{1}, S^{2},S^{3})$:
\begin{eqnarray}
\xi = 2N^{2} \; , \qquad \eta = 2M^{2} \; , \qquad \phi = \Delta
\; . \nonumber
\\
x = S^{1} \; , \qquad y = S^{2} \; , \qquad z = S^{3} \; .
\end{eqnarray}

 Let us  find a space spinor
$\Psi_{space}$ related to space coordinates $(x,y,z)$ on the base
of relationship  (see Cartan \cite{1938-Cartan})
\begin{eqnarray}
\Psi_{space} = \left | \begin{array}{r} N \;e^{i\alpha}  \\ M\;
e^{i\beta}
\end{array} \right | , \qquad \psi \otimes \psi^{*}=
  {1 \over 2} \left | \begin{array}{cc}
r + z & x - i y     \\
 x + i y           &  r -z
 \end{array} \right |\; ,
\label{4.19'}
\end{eqnarray}

\noindent one  would produce (for more detail see
\cite{2004-Redkov}, \cite{2005-Redkov}) the formulas  like (\ref{4.24}):
\begin{eqnarray}
\Psi_{space}  = e^{-\gamma /2} \;  \left | \begin{array}{c}
\sqrt{r + z } \;  e^{-i\phi/2}  \\
\sqrt{r - z} \;  e^{+i\phi /2}  \end{array} \right | \nonumber \\
= e^{-\gamma /2} \;  \left | \begin{array}{c}
\sqrt{\xi} \;  e^{-i\phi/2}  \\
\sqrt{\eta} \;  e^{+i\phi /2}  \end{array} \right |
 , \;\;
e^{i\phi } = { x + i y \over \sqrt{ x^{2} + y^{2} } } \; .
\label{A.6'}
\end{eqnarray}

Spinor $\Psi$ (or $\Psi_{space}$)   has evident  peculiarity:
  at the  whole axis  $S_{1}=S_{2}=0$  (or  at $x = y =0$) its defining relation  contains ambiguity
$(0+i0)/0$ (and expressions for $\xi$ will contain a mute  angle
variable $\Gamma: \;
 \phi  \rightarrow \Gamma$ )
\begin{eqnarray}
(0,0, S_{3} > 0) \; : \qquad \Psi^{+} _{0}= \left |
\begin{array}{c} \sqrt{+2S_{3}} \; e^{-i\Delta /2} \\ 0
\end{array} \right | \; , \nonumber
\\
(0,0, S_{3} < 0) \;\; :  \qquad
 \Psi^{-}_{0}  = \left | \begin{array}{c}
 0  \\ \sqrt{-2S_{3} }\; e^{+i\Delta /2} \end{array} \right | \;,
\nonumber
\\
e^{i\Gamma }  =  \lim_{S_{1},S_{2} \rightarrow 0} { S_{1} + i
S_{2} \over \sqrt{ S^{2}_{1} + S^{2}_{2} }}\; . \label{A.7c'}
\qquad \qquad
\end{eqnarray}

It should be mention that polarization singularities, attracting
attention in
 the literature \cite{2008-Bliokh-Niv-Kleiner-Hasman},
should be associated with that peculiarities $(0+i0)/0$  in
(\ref{A.7c'}); in other words, they are peculiarities in parameterizing
space spinor $\Psi_{space}$  by parabolic coordinates $(\xi, \eta,
\phi)$.

Also one can take special attention to the  factors $e^{+i\phi /2}
$ and  $e^{-i\phi /2} $ in expression for Jones spinor, which
leads to ($\pm$)-ambiguity at the values $\Phi = 0$ and $\Phi = +
2\pi$ or $\Delta = 0$ and $\Delta = + 2\pi$. However these two
values correspond physically to one the same direction in
geometrical space or to one the same Stokes 3-vector.  It is an
old problem with spinors applied to description of 3-vectors, and
it can be overcome in the frame ideas on spinor space structure:
\cite{1984-Penrose}; also see in \cite{2004-Redkov},
\cite{2005-Redkov} and  references therein.
 Does the spinor group topology
is relevant indeed to Jones complex formalism in optics or not --
the issue remains open, for theory and experiments.

Also, we might  take special attention to the fact,  that usually
the space vector $(x,y,z)$ is not taken to be a pseudovector,  but
constructing $(x,y,z)$  in spinor approach   according to
(\ref{4.19'}) leads just to a such pseudo-vector.

Let us recall Cartan classification for non-relativistic spinors
 \cite{1938-Cartan}
of 2-spinors with respect to spinor $P$-reflection: namely,  the
simplest irreducible representations of the  unitary extended
group
\begin{eqnarray}
\tilde{SU}(2) =  \Big\{\;\; g \in SU(2) \oplus J = \left |
\begin{array}{cc} i & 0 \\ 0  & i \end{array} \right | , \;\;
   \det \; g = +1, \; \det \; J = -1 \;\;  \Big\} \qquad
\nonumber
\end{eqnarray}

\noindent are 2-component spinors of two types $T_{1},T_{2}$
\begin{eqnarray}
  T_{1}(g) = g \; , \;\;  T_{1}(J) = + J \; ; \qquad \qquad
 T_{2}(g) = g\; , \; \; T_{2}(J) = - J \; .
\nonumber
\end{eqnarray}

\noindent There are two ways to
 construct 3-vector (complex-valued in general)
in terms of 2-spinors
\begin{eqnarray}
1. \qquad (\Psi_{space} \otimes \Psi_{space} ^{*}) =
 r \;+\; x_{j} \; \sigma ^{j}\; \; , \;\;
r  = + \sqrt{ x_{j} \; x_{j}} \; , \qquad
x_{j}-\mbox{pseudovector}\; ; \nonumber
\\
2.  \qquad \qquad \qquad \;\;\; (\Psi'_{space}\otimes
\Psi'_{space} ) =  \;( y_{j} \;+\; i \; x_{j}) \;\sigma
^{j}\;\sigma^{2} \; , \qquad y_{j},x_{j}-\mbox{vectors}\; .
\label{A.3'}
\end{eqnarray}

Evidently, variant 1 provides us  with  a possibility to build a
spinor model for  pseudo vector 3-space, whereas variant 2 leads
to a spinor model of  properly vector 3-space. In other words,
according to different  ways of taking the square root of three
real numbers -- components  of a 3-vector $x_{i}$ -- one
arrives at two different spatial spinors:
\begin{eqnarray}
\Psi_{space} \;  \Longleftrightarrow  \; x_{j} \; , \qquad
\Psi'_{space}\;  \Longleftrightarrow  \; x_{j} \; . \nonumber
\end{eqnarray}

These spinors, $\Psi_{space}$ and $\Psi'_{space}$ respectively,
turned out to be  different functions of Cartesian coordinates. In
particular, the second spinor model corresponding to  a vector
space, variant 2. in  (\ref{A.3'}), is  described by two spinors
$\Psi'_{space}( \vec{x})$, each   covering a vector  half-space
(for more detail see \cite{2004-Redkov}, \cite{2005-Redkov}):
\begin{eqnarray}
 x_{3} > 0 \; , \qquad
\Psi_{space} ^{'+} = \left | \begin{array}{c}
\sqrt{r - (x^{2} + y^{2})^{1/2} } \; e^{-i\phi /2} \\[2mm]
\sqrt{r + (x^{2} + y^{2})^{1/2} } \; e^{+i\phi /2}
\end{array} \right |
 \; , \qquad
\; e^{i\phi } = { x + i y \over \sqrt{ x^{2} + y^{2} }} \;.
\nonumber
\end{eqnarray}
\begin{eqnarray}
 x_{3} < 0 \; , \qquad
\Psi_{space}^{'-} = i  \left | \begin{array}{c}
\sqrt{ r - (x^{2} + y^{2})^{1/2}} \; e^{-i\sigma /2} \\[2mm]
\sqrt{ r + (x^{2} + y^{2})^{1/2}} \; e^{+i\sigma /2}
\end{array} \right |  \; ,  \qquad
e^{i\sigma /2}  = - i \sqrt{{x + i y} \over \sqrt{x^{2} + y^{2}}
}\; . \nonumber
\\
\label{A.6b'}
\end{eqnarray}

In the context of polarization optics instead of (\ref{A.6b'}) we
would have
\begin{eqnarray}
 S^{3} > 0 \;  , \qquad
\Psi'_{+} = \left | \begin{array}{c}
N' e^{-i\Delta /2} \\
M'  e^{+i\Delta /2}
\end{array} \right | =
\left | \begin{array}{c}
\sqrt{S - (S^{2}_{1} + S^{2}_{2})^{1/2} }  e^{-i\Delta /2} \\[2mm]
\sqrt{S + (S^{2}_{1} + S^{2}_{2})^{1/2} }  e^{+i\Delta /2}
\end{array} \right |
  , \;
\; e^{i\Delta } = { S^{1}+ i S^{2} \over \sqrt{ S^{2}_{1} +
S^{2}_{2} }} . \nonumber
\end{eqnarray}
\begin{eqnarray}
 S^{3} < 0 \;  , \qquad
\Psi'_{-} =\left | \begin{array}{c}
N' e^{-i\Delta /2} \\
M '  e^{+i\Delta /2}
\end{array} \right |
=  i  \left | \begin{array}{c}
\sqrt{ S - (S^{2}_{1} + S^{2}_{2})^{1/2}} \; e^{-i\sigma /2} \\[2mm]
\sqrt{ S + (S^{2}_{1} + S^{2}_{2})^{1/2}} \; e^{+i\sigma /2}
\end{array} \right |   ,  \;
e^{i\sigma /2}  = - i \sqrt{{S^{1} + i S^{2} \over \sqrt{S^{2}_{1}
+ S^{2}_{2}}} } . \nonumber
\\
\label{A.6b''}
\end{eqnarray}

\noindent Here again we have singularity in parametrization on the
whole axis $x_{1}=0,x_{2}=0$ or at $S^{1} =0, \; S^{2} = 0 $.

Two models of spinors spaces  with respect to $P$-orientation are
grounded  on different mappings  $\Psi_{\space} ({\bf x})$   and
$\Psi'_{space} ({\bf x})  $  defined over  the same extended
domain $\tilde{G}(x_{i})$.  The natural question is:  how are
these two  maps  connected to each others. An answer  can be
found on comparing the formulas
\begin{eqnarray}
\Psi_{space} = \left | \begin{array}{c} \sqrt{\xi}  \; e^{-i \phi
/2  }   \\   \sqrt{\eta} \; e^{+i \phi /2}
\end{array} \right | \; ,
 \qquad
\Psi'_{space}  = {1 \over \sqrt{2}} \left | \begin{array}{c}
(\sqrt{\xi} - \sqrt{\eta}) \; e^{-i \phi/2}  \\
(\sqrt{\xi} + \sqrt{\eta}) \; e^{+i \phi/2}
\end{array} \right | \;  .
\label{b4.1}
\end{eqnarray}

\noindent From (\ref{b4.1}) we   immediately arrive at
\begin{eqnarray}
\Psi'_{space}  = {1 \over \sqrt{2}} ( \Psi_{space} - i \; \sigma
^{2} \Psi ^{*} _{space}) \; , \nonumber
\\
\Psi_{space} = {1 \over \sqrt{2}} \; ( \Psi'_{space} -  i \;\sigma
^{2} \Psi_{space}^{'*} ) \; . \label{b4.3}
\end{eqnarray}

Finally, let us write  down the formulas for Stoke 3-vector in
both  cases:

\vspace{3mm}  traditional model $\Psi ({\bf S})$
\begin{eqnarray}
S^{1} = \sqrt{{N M \over 2}} \; \cos \Delta \; , \qquad S^{2} =
\sqrt{{N M \over 2}} \; \sin \Delta \; , \qquad S^{3} = N^{2} -
M^{2} \; ;
\label{end'}
\end{eqnarray}

\vspace{3mm}  alternative  model $\Psi' ({\bf S})$
\begin{eqnarray}
S^{1} = \sqrt{2 \mid M^{'2} - N^{'2} \mid }  \cos \Delta \; ,\;
S^{2} = \sqrt{2 \mid M^{'2} - N^{'2} \mid }  \sin \Delta \; ,
\qquad S^{3} = \pm \; \sqrt{N'M'} \; . \label{end''}
\end{eqnarray}

\section{ Spinor representation of  Stokes 4-vector \\ and  2-rank tensor for a  completely  polarized
light}

Let start with the well-known relations between  2-rank bi-spinors
and simplest tensors. Bi-spinor of second rank $U  = \Psi \otimes
\Psi $ can be  resolved into scalar  $\Phi $, vector $\Phi _{b}$ ;
pseudoscalar $\tilde{\Phi } $, pseudovector  $\tilde{\Phi }_{b} $,
and antisymmetric tensor   $\Phi _{ab}$\footnote{In this section
we use  Dirac matrices, for more details  see in   \cite{Book}.}
\begin{eqnarray}
U  = \Psi \otimes \Psi  = \left [\;  - i \; \Phi  + \gamma^{b} \;
\Phi _{b}  +
           i \; \sigma^{ab}\;  \Phi _{ab}  +  \gamma ^{5} \; \tilde{\Phi }  +
           i \; \gamma ^{b} \gamma ^{5} \; \tilde {\Phi }_{b}  \; \right ]\; E^{-1} \; ;
\label{9.5.1}
\end{eqnarray}

\noindent let us refer all consideration to the  spinor basis
\begin{eqnarray}
U  =
\left | \begin{array}{cc}
\xi ^{\alpha \beta }  & \Delta ^{\alpha}_{\;\;\; \dot{\beta }} \\
H_{\dot{\alpha}} ^{\;\;\;\beta}  & \eta _{\dot{\alpha}\dot{\beta
}}
\end{array} \right | \; , \qquad
\gamma^{a} = \left | \begin{array}{cc}
0 & \bar{\sigma}^{a} \\
\sigma^{a} & 0   \end{array} \right | \; , \;
\gamma^{5}  =
\left | \begin{array}{cc} - I & 0 \\ 0 & +I \end{array} \right |
\;,
\nonumber
\\
\sigma^{ab}  = {1 \over 4}  \; \left | \begin{array}{cc}
\bar{\sigma}^{a} \sigma^{b} - \bar{\sigma}^{b} \sigma^{a} & 0 \\
0  & \sigma^{a} \bar{\sigma}^{b} - \sigma^{b} \bar{\sigma}^{a}
\end{array} \right | =
\left | \begin{array}{cc} \Sigma^{ab} & 0 \\ 0  &
\bar{\Sigma}^{ab}  \end{array} \right | \;  ; \label{9.5.4}
\end{eqnarray}

\noindent  $E$ stands for a bi-spinor metric matrix
\begin{eqnarray}
E =  \left | \begin{array}{cc}
               \epsilon   &   0  \\ 0   &   \dot{\epsilon}^{-1}
\end{array}  \right | =
     \left | \begin{array}{cc}
               \epsilon_{\alpha \beta}  &   0  \\
                0      &      \epsilon ^{\dot{\alpha}\dot{\beta}}
     \end{array}  \right | =
     \left | \begin{array}{cc}
              i  \sigma^{2}  &   0  \\
                0      &    - i   \sigma^{2}
     \end{array}  \right |  .
\nonumber
\end{eqnarray}

\noindent Inverse to (\ref{9.5.1}) relations look
\begin{eqnarray}
\Phi _{a}  = {1\over 4} \; \mbox{Sp}  \; [ E \gamma _{a} U ] \; ,
\qquad \tilde{\Phi }_{a}  = {1\over 4i}  \; \mbox{Sp} \; [E \gamma
^{5}\gamma _{a} U ] \; , \nonumber
\\
\Phi  = {i \over 4} \; \mbox{Sp} \; [ E U  ] \; , \qquad
\tilde{\Phi }  = {1\over 4} \;  \mbox{Sp}\;  [E \gamma ^{5} U ] \;
, \;
\Phi _{mn} = -{1 \over 2i} \; \mbox{Sp} \;  [E \sigma _{mn} U ]\;
. \label{9.5.3}
\end{eqnarray}

First, we are interested in two vectors obtained from spinors:
\begin{eqnarray}
\Phi_{a} =  {1 \over 2 } \mbox{Sp} \left | \begin{array}{cc}
i \sigma^{2} \bar{\sigma}_{a} H &  i \sigma^{2} \bar{\sigma}_{a} \eta \\
-i \sigma^{2} \sigma_{a} \xi & - i \sigma^{2} \sigma_{a} \Delta
\end{array} \right |,
\nonumber
\end{eqnarray}

\noindent so that
\begin{eqnarray}
\Phi_{0} =   {1 \over 2} \; [ \; (H_{\dot{2}}^{\;\; 1} -  H_{\dot{1}}^{\;\; 2} ) -
( \Delta^{2}_{\;\;\dot{1}}  -  \Delta^{1}_{\;\;\dot{2}} ) \; ] =
\xi^{1}  \eta_{\dot{2}}  - \xi^{2}  \eta_{\dot{1}} \; , \nonumber
\\
\Phi_{1} =   {1 \over 2}\; [ \; (H_{\dot{1}}^{\;\;1} -  H_{\dot{2}}^{\;\;2}) +
(\Delta^{1}_{\;\;\dot{1}} -  \Delta^{2}_{\;\;\dot{2}} )\; ] =
\xi^{1} \eta_{\dot{1}}  - \xi^{2} \eta_{\dot{2}}  \; , \nonumber
\\
\Phi_{2} =
 { i \over 2}\; [  \; ( H_{\dot{1}}^{\;\;1} + H_{\dot{2}}^{\;\;2} ) +
 (\Delta^{1}_{\;\;\dot{1}} + \Delta^{2}_{\;\;\dot{2}}  ) \; ] =
   i \;  (  \xi^{1} \eta_{\dot{1}} + \xi^{2} \eta_{\dot{2}}  ) \; ,
\nonumber
\\
\Phi_{3} =  -{ 1 \over 2}\; [ \;  ( H_{\dot{2}}^{\;\;1} + H_{\dot{1}}^{\;\;2} ) +
 (\Delta^{2}_{\;\;\dot{1}} + \Delta^{1}_{\;\;\dot{2}}  ) \; ] =
   - \;  (  \xi^{1} \eta_{\dot{2}} + \xi^{2} \eta_{\dot{1}}  )\; ;
\nonumber
\end{eqnarray}

\noindent and for pseudovector
\begin{eqnarray}
\tilde{\Phi}_{a} =  {1 \over 2 }   \mbox{Sp} \left |
\begin{array}{cc}
-i \sigma^{2} \bar{\sigma}_{a} H &  - i \sigma^{2} \bar{\sigma}_{a} \eta \\
-i \sigma^{2} \sigma_{a} \xi & - i \sigma^{2} \sigma_{a} \Delta
\end{array} \right | \; ,
\nonumber
\end{eqnarray}

\noindent so that
\begin{eqnarray}
\tilde{\Phi}_{0} =  {1 \over 2} \; [ \; -(H_{\dot{2}}^{\;\; 1} -  H_{\dot{1}}^{\;\; 2} ) -
( \Delta^{2}_{\;\;\dot{1}}  -  \Delta^{1}_{\;\;\dot{2}} ) \; ] =
0 \; , \nonumber
\\
\tilde{\Phi}_{1} =  {1 \over 2}\; [ \;- (H_{\dot{1}}^{\;\;1} -  H_{\dot{2}}^{\;\;2}) +
(\Delta^{1}_{\;\;\dot{1}} -  \Delta^{2}_{\;\;\dot{2}} )\; ] = 0
\; , \nonumber
\\
\tilde{\Phi}_{2} =
 { i \over 2}\; [  \; - ( H_{\dot{1}}^{\;\;1} + H_{\dot{2}}^{\;\;2} ) +
 (\Delta^{1}_{\;\;\dot{1}} + \Delta^{2}_{\;\;\dot{2}}  ) \; ] = 0
\; , \nonumber
\\
\tilde{\Phi}_{3} =  -{ 1 \over 2}\; [ \; - ( H_{\dot{2}}^{\;\;1} + H_{\dot{1}}^{\;\;2} ) +
 (\Delta^{2}_{\;\;\dot{1}} + \Delta^{1}_{\;\;\dot{2}}  ) \; ] = 0 \; .
\nonumber
\end{eqnarray}

\noindent In the same manner we get for scalar and pseudoscalar:
\begin{eqnarray}
\Phi =  {i\over 4} \;
[\; + (\xi^{21} - \xi^{12} ) - ( \eta_{\dot{2}\dot{1}} +
\eta_{\dot{1}\dot{2}})\; ] = 0 \; ,
\nonumber
\\
\tilde{\Phi} =  {i\over 4} \;
[\;- (\xi^{21} - \xi^{12} ) - ( \eta_{\dot{2}\dot{1}} +
\eta_{\dot{1}\dot{2}})\; ] = 0 \; ; \nonumber
\end{eqnarray}

\noindent and for antisymmetric tensor
\begin{eqnarray}
\Phi ^{mn} = - {1 \over 2i} \; \mbox{Sp} \;  [ \; E \sigma ^{mn} U
\; ] = -{1 \over 2i} \; \mbox{Sp} \; \left | \begin{array}{cc}
i \sigma^{2} \Sigma^{mn} \xi & ... \\
... & - i \sigma^{2} \bar{\Sigma}^{mn} \eta
\end{array} \right | \; ,
\nonumber
\end{eqnarray}

\noindent so that
\begin{eqnarray}
\Phi^{01} =  {i \over 4} \;   [ \; (\xi^{11} - \xi^{22}) + (
\eta_{\dot{1}\dot{1}} -  \eta_{\dot{2}\dot{2}}) \;  ]= {i \over 4}
\;  [ \; (\xi^{1} \xi^{1} - \xi^{2} \xi^{2}) + ( \eta_{\dot{1}}
\eta_{\dot{1}} -  \eta_{\dot{2}} \eta_{\dot{2}}) \;  ]   \; ,
\nonumber
\\
\Phi^{23} =  {1 \over 4}  \;  [ \;  (\xi^{11} - \xi^{22}) - (
\eta_{\dot{1}\dot{1}} -  \eta_{\dot{2}\dot{2}})  \; ]= {1 \over 4}
\;   [\;  (\xi^{1} \xi^{1} - \xi^{2} \xi^{2}) - ( \eta_{\dot{1}}
\eta_{\dot{1}} -  \eta_{\dot{2}} \eta_{\dot{2}}) \;  ]   \; ,
\nonumber
\\[2mm]
\Phi^{02} = - {1 \over 4} \;   [  \; (\xi^{11} + \xi^{22}) + (
\eta_{\dot{1}\dot{1}} +  \eta_{\dot{2}\dot{2}})  \; ] = - {1 \over
4}   \; [  \; (\xi^{1} \xi^{1} + \xi^{2} \xi^{2}) + (
\eta_{\dot{1}} \eta_{\dot{1}} +  \eta_{\dot{2}} \eta_{\dot{2}})\;
]\; , \nonumber
\\
\Phi^{31} =   - {1 \over 4i} \;   [  \; (\xi^{11} + \xi^{22}) - (
\eta_{\dot{1}\dot{1}} +  \eta_{\dot{2}\dot{2}})  \; ] = - {1 \over
4i}   \; [  \; (\xi^{1} \xi^{1} + \xi^{2} \xi^{2}) - (
\eta_{\dot{1}} \eta_{\dot{1}} +  \eta_{\dot{2}} \eta_{\dot{2}})\;
]\; , \nonumber
\\[2mm]
\Phi^{03} = - {i \over 4}  \;  [ \;  (\xi^{21} + \xi^{12}) + (
\eta_{\dot{2}\dot{1}} +  \eta_{\dot{1}\dot{2}})  ]= - {i \over 2 }
\;  [  \;  \xi^{1} \xi^{2} + \eta_{\dot{1}} \eta_{\dot{2}}  ] \; ,
\nonumber
\\
\Phi^{12} =  - {1 \over 4}  \;  [ \;  (\xi^{21} + \xi^{12}) - (
\eta_{\dot{2}\dot{1}} +  \eta_{\dot{1}\dot{2}})  ]= - {1 \over 2 }
\;  [  \;  \xi^{1} \xi^{2} - \eta_{\dot{1}} \eta_{\dot{2}}  ] \; ,
\nonumber
\end{eqnarray}

Collecting results together:
\begin{eqnarray}
\Psi = \left | \begin{array}{c}
\xi^{\alpha} \\
\eta_{\dot{\alpha}}
\end{array} \right |\; , \qquad \Psi \otimes \Psi  \qquad  \Longrightarrow \qquad
\nonumber
\\
\Phi =0 , \;\; \tilde{\Phi} = 0, \;\; \tilde{\Phi}_{a} = 0 , \;\;
{\Phi}_{a} \neq 0 ,\;\;  {\Phi}_{mn} \neq 0 \; , \nonumber
\end{eqnarray}

\noindent we see that to have real vector  and tensor one should
impose additional restriction: for example let it be
\begin{eqnarray}
\eta = + i \; \sigma^{2}  \; \xi^{*} \; \qquad \Longrightarrow
\qquad \eta_{\dot{1}} = + \; \xi^{2*}\; , \;\; \eta_{\dot{2}} =  -
\; \xi^{1*} \; ; \label{reality}
\end{eqnarray}

\noindent which results in
\begin{eqnarray}
\Phi_{0} =   - (\xi^{1} \;  \xi^{1*}  + \xi^{2}  \;  \xi^{2*})  <
0 \; , \qquad \Phi_{3} =     (\xi^{1}  \; \xi^{1*} - \xi^{2}  \;
\xi^{2*}  ) \; ,
\nonumber
\\
\Phi_{1} = ( \xi^{1}  \; \xi^{2*}   + \xi^{2}  \; \xi^{1*} )\; ,
\qquad \Phi_{2} =
    i \;  (  \xi^{1}  \; \xi^{2*}  - \xi^{2} \; \xi^{1*}  )\; ;
\nonumber
\\[2mm]
\Phi^{01} = {i \over 4}  \;   [\;  (\xi^{1} \; \xi^{1} - \xi^{2}
\; \xi^{2}) + (  \xi^{2*} \; \xi^{2*}  -  \xi^{1*} \; \xi^{1*} )
\;  ]
 \; ,
\nonumber
\\
\Phi^{23} = {1 \over 4} \;   [\;  (\xi^{1}  \; \xi^{1} - \xi^{2}
\;  \xi^{2}) - (  \xi^{2*} \; \xi^{2*}  -  \xi^{1*} \; \xi^{1*} )
\;  ]   \; , \nonumber
\\
\Phi^{02} = - {1 \over 4}   \; [  \; ( \xi^{1} \; \xi^{1} +
\xi^{2} \; \xi^{2}) + (  \xi^{2*} \; \xi^{2*}  +  \xi^{1*} \;
\xi^{1*}  )\;   ]\; , \nonumber
\\[2mm]
\Phi^{31} =  - {1 \over 4i}   \; [  \; (\xi^{1} \;  \xi^{1} +
\xi^{2} \; \xi^{2}) - ( \xi^{2*} \; \xi^{2*}  +   \xi^{1*} \;
\xi^{1*} )\;   ] \; , \nonumber
\\
\Phi^{03} =  - {i \over 2 }  \;  (  \;  \xi^{1}  \; \xi^{2} -
\xi^{2*}  \;  \xi^{1*}  ) \; , \qquad \Phi^{12} = - {1 \over 2 }
\;  [  \;  \xi^{1} \; \xi^{2} + \xi^{2*} \; \xi^{1*}  ] \; ,
\label{real}
\end{eqnarray}

There exists alternative additional restriction:
\begin{eqnarray}
\eta = - i \; \sigma^{2}  \; \xi^{*} \; \qquad \Longrightarrow
\qquad \eta_{\dot{1}} = - \xi^{2*}\; , \;\; \eta_{\dot{2}} =  +
\xi^{1*} \; , \label{reality'}
\end{eqnarray}

\noindent which results in (compare with (\ref{real}))

\begin{eqnarray}
\Phi_{0} =   ( \xi^{1} \;  \xi^{1*}  + \xi^{2}  \;  \xi^{2*}  ) >
0  \; , \qquad \Phi_{3} =     - ( \xi^{1}  \; \xi^{1*} - \xi^{2}
\; \xi^{2*})    \; ,
\nonumber
\\
\Phi_{1} =- (
 \xi^{1}  \; \xi^{2*}   + \xi^{2}  \; \xi^{1*}) \; , \qquad
\Phi_{2} =
   - i \;  (  \xi^{1}  \; \xi^{2*}  - \xi^{2} \; \xi^{1*}  )\; ;
\nonumber
\\[2mm]
\Phi^{01} = {i \over 4}  \;   [\;  (\xi^{1} \; \xi^{1} - \xi^{2}
\; \xi^{2}) + (  \xi^{2*} \; \xi^{2*}  -  \xi^{1*} \; \xi^{1*} )
\;  ]
 \; ,
\nonumber
\\
\Phi^{23} = {1 \over 4} \;   [\;  (\xi^{1}  \; \xi^{1} - \xi^{2}
\;  \xi^{2}) - (  \xi^{2*} \; \xi^{2*}  -  \xi^{1*} \; \xi^{1*} )
\;  ]   \; , \nonumber
\\[2mm]
\Phi^{02} = - {1 \over 4}   \; [  \; ( \xi^{1} \; \xi^{1} +
\xi^{2} \; \xi^{2}) + (  \xi^{2*} \; \xi^{2*}  +  \xi^{1*} \;
\xi^{1*}  )\;   ]\; , \nonumber
\\
\Phi^{31} =  - {1 \over 4i}   \; [  \; (\xi^{1} \;  \xi^{1} +
\xi^{2} \; \xi^{2}) - ( \xi^{2*} \; \xi^{2*}  +   \xi^{1*} \;
\xi^{1*} )\;   ] \; , \nonumber
\\[2mm]
\Phi^{03} =  - {i \over 2 }  \;  (  \;  \xi^{1}  \; \xi^{2} -
\xi^{2*}  \;  \xi^{1*}  ) \; , \qquad \Phi^{12} = - {1 \over 2 }
\;  [  \;  \xi^{1} \; \xi^{2} + \xi^{2*} \; \xi^{1*}  ] \; ,
\label{real'}
\end{eqnarray}

The last case  (\ref{reality'}) -- (\ref{reality'}) seems to be
appropriate  to describe Stokes 4-vector and  determine Stokes
2-rank tensor:

\begin{eqnarray}
\Psi = \left | \begin{array}{c}
\xi \\
\eta = - i \; \sigma^{2}  \; \xi^{*}
\end{array} \right |\; , \qquad \Psi \otimes \Psi  \qquad  \Longrightarrow \qquad
 S_{a} \neq 0 ,\;\;  S_{mn} \neq 0 \; ,
\nonumber
\\[2mm]
S_{0} =   ( \xi^{1} \;  \xi^{1*}  + \xi^{2}  \;  \xi^{2*}  ) > 0
\; , \qquad S_{3} =     - ( \xi^{1}  \; \xi^{1*} - \xi^{2}  \;
\xi^{2*})    \; ,
\nonumber
\\
S_{1} =- ( \xi^{1}  \; \xi^{2*}   + \xi^{2}  \; \xi^{1*}) \; ,
\qquad S_{2} =
   - i \;  (  \xi^{1}  \; \xi^{2*}  - \xi^{2} \; \xi^{1*}  )\; ;
\nonumber
\\[2mm]
a^{1} = S^{01} = {i \over 4}  \;   [\;  (\xi^{1} \; \xi^{1} -
\xi^{2} \; \xi^{2}) + (  \xi^{2*} \; \xi^{2*}  -  \xi^{1*} \;
\xi^{1*} ) \;  ]
 \; ,
\nonumber
\\
b^{1} = S^{23} = {1 \over 4} \;   [\;  (\xi^{1}  \; \xi^{1} -
\xi^{2} \;  \xi^{2}) - (  \xi^{2*} \; \xi^{2*}  -  \xi^{1*} \;
\xi^{1*} ) \;  ]   \; ,
\nonumber
\\[2mm]
a^{2} = S^{02} = - {1 \over 4}   \; [  \; ( \xi^{1} \; \xi^{1} +
\xi^{2} \; \xi^{2}) + (  \xi^{2*} \; \xi^{2*}  +  \xi^{1*} \;
\xi^{1*}  )\;   ]\; , \nonumber
\\
b^{2} = S^{31} =  - {1 \over 4i}   \; [  \; (\xi^{1} \;  \xi^{1} +
\xi^{2} \; \xi^{2}) - ( \xi^{2*} \; \xi^{2*}  +   \xi^{1*} \;
\xi^{1*} )\;   ] \; ,
\nonumber
\\[2mm]
a^{3} = S^{03} =  - {i \over 2 }  \;  (  \;  \xi^{1}  \; \xi^{2} -
\xi^{2*}  \;  \xi^{1*}  ) \; , \qquad b^{3} = S^{12} = - {1 \over
2 }  \;  (  \;  \xi^{1} \; \xi^{2} + \xi^{2*} \; \xi^{1*}  ) \; .
\label{newStokes}
\end{eqnarray}

Let us calculate the main invariant -- it turns to equal to zero:
\begin{eqnarray}
S_{0}S_{0} - S_{j}S_{j} =  0 \; ,
\end{eqnarray}

\noindent so $S_{a}$ may be considered as a Stokes 4-vector for a
completely polarized light.

In turn, 4-tensor $S_{mn}$, being constructed
from Jones bi-spinor $\Psi$,  is a Stokes 2-rank tensor.
Let us calculate two invariants for $S_{mn}$:
\begin{eqnarray}
I_{1}= - {1 \over 2}\; S^{mn}S_{mn} = {\bf a}^{2} - {\bf b}^{2} = 0\; , \qquad
I_{2} = {1 \over 4}\; \epsilon_{abmn} S^{ab} S^{mn}  = 0 \; .
\label{inv-s}
\end{eqnarray}

Finally, let us  specify Stokes 4-vector and 4-tensor in parameters
 $(M,N,\Delta = \alpha - \beta)$ :
\begin{eqnarray}
\Psi = \left | \begin{array}{r}
N \; e^{i\alpha } \\ + M \;  e^{i \beta}  \\ - M \; e^{-i \beta}\\
N \;  e^{-i\alpha }
\end{array} \right |
\; , \qquad \Psi \otimes \Psi  \qquad  \Longrightarrow \qquad
 S_{a} \neq 0 ,\;\;  S_{mn} \neq 0 \; ,
\nonumber
\\
S_{0} =  M^{2}  +  N^{2}   \; , \qquad S_{3} =  M^{2}   -  N^{2}
\; , \nonumber
\\
S_{1} =- 2MN \cos (\alpha - \beta)  \; , \qquad S_{2} = 2MN \sin
(\alpha - \beta)
 \;
\nonumber
\end{eqnarray}

\noindent which coincides with (\ref{4.21}); and
\begin{eqnarray}
a^{1} = S^{01} = -{1 \over 2} (N^{2} \sin 2\alpha - M^{2} \sin 2
\beta ) \; , \;
b^{1} = S^{23} = + {1 \over 2} (N^{2} \cos 2\alpha - M^{2} \cos 2
\beta ) \; , \nonumber
\\
a^{2} = S^{02} =
 -{1 \over 2} (N^{2} \cos 2\alpha + M^{2} \cos 2 \beta ) \; ,
\;
b^{2} = S^{31} =
 -{1 \over 2} (N^{2} \sin 2\alpha + M^{2} \sin 2 \beta ) \; ,
\nonumber
\\
a^{3} = S^{03} =  + NM \sin (\alpha + \beta) \; , \qquad
b^{3} = S^{12} = -  NM \cos ( \alpha  + \beta ) \; .
\label{newStokes'}
\end{eqnarray}

Two vectors ${\bf a}, {\bf b}$ are determined by 4 parameters $
N,M, \alpha, \beta $, additional  identities hold
\begin{eqnarray}
{\bf a}^{2} = {\bf b}^{2} = {(N^{2} + M^{2})^{2} \over 4   } \; ,
\qquad {\bf a} {\bf b} = 0 \; ; \nonumber
\end{eqnarray}

\noindent therefore the quantities ${\bf a}, {\bf b}$ depend in
fact upon 4 independent parameters $N,M, \beta - \alpha, \beta +
\alpha $; whereas Stokes 4-vector depends upon only three ones
$N,M, \beta - \alpha$.

Instead of Stokes 4-tensor $S_{ab}$ one may introduce a complex
Stokes 3-vector  ${\bf s}= {\bf a} + i {\bf b}$ with the
 components (see (\ref{newStokes})):
\begin{eqnarray}
s^{1} = a^{1}  + i b^{1} = S^{01} + i S^{23} = {i \over 2}  \;
(\xi^{1} \; \xi^{1} - \xi^{2} \; \xi^{2}) \; , \nonumber
\\
s^{2} = a^{2} + i b^{2} =  S^{02} + i S^{31}  = - {1 \over 2}   \;
( \xi^{1} \; \xi^{1} + \xi^{2} \; \xi^{2}) \; , \nonumber
\\
s^{3} = a^{3} + i b^{3} = S^{03} = i S^{12} =  - i   \;    \xi^{1}
\; \xi^{2}    \; ; \label{complexStokes}
\end{eqnarray}

\noindent from whence it follows
\begin{eqnarray}
s_{1} + i s_{2} = - i\; \xi^{2} \xi^{2} \; , \qquad s_{1} - i
s_{2} = + i\; \xi^{1} \xi^{1} \; , \qquad s^{3}   - i   \;    \;
\xi^{1}  \; \xi^{2}    \; .
\end{eqnarray}

The quantity ${\bf s}$  transforms as a vector under complex
rotation group $SO(3.C)$, isomorphic to Lorentz group $L^{\uparrow}_{+}$. The later permits to introduce
additionally to  Jones spinor and Mueller vector formalisms one
other technique based on the use of complex 3-vector under complex
rotation group $SO(3.C)$:
\begin{eqnarray}
{\bf s} = {\bf a} + i {\bf b} = {1 \over 2} \left |
\begin{array}{r} i\; ( N^{2} e^{2i\alpha} - M^{2} e^{2i \beta} )
\\
- \;(N^{2} e^{2i\alpha} + M^{2} e^{2i \beta}  ) \\
-2i\; NM \;  e^{i(\alpha + \beta) }
\end{array} \right | ;
\end{eqnarray}

\noindent evidently this complex vector is isotropic ${\bf s}^{2}
= 0 $, the later condition provide us with two additional
condition, so ${\bf s}$ depends on 4 parameters.

\section{ Spinor representation for a space-time vectors}

The question relevant to description of a  partly polarized  in
terms of Jones spinor object still remains unsolved. Let us turn
back and consider possibility to construct vector and tensor in
terms of spinor components
 with no additional restriction  on bi-spinor (like   $\eta = \pm  i \sigma^{2} \xi^{*}$):
\begin{eqnarray}
\Phi_{0} =  \xi^{1}  \eta_{\dot{2}}  - \xi^{2}  \eta_{\dot{1}} \;
, \qquad \Phi_{3} =   - \;  (  \xi^{1} \eta_{\dot{2}} + \xi^{2}
\eta_{\dot{1}}  )\; , \nonumber
\\
\Phi_{1} =   \xi^{1} \eta_{\dot{1}}  - \xi^{2} \eta_{\dot{2}}  \;
, \qquad \Phi_{2} =
   i \;  (  \xi^{1} \eta_{\dot{1}} + \xi^{2} \eta_{\dot{2}}  ) \; ;
\nonumber
\\[2mm]
\Phi_{0}^{2} - \Phi_{3}^{2} = (\xi^{1}  \eta_{\dot{2}}  - \xi^{2}
\eta_{\dot{1}})^{2} - (\xi^{1} \eta_{\dot{2}} + \xi^{2}
\eta_{\dot{1}})^{2} = - 4  \; \xi^{1} \xi^{2} \; \eta_{\dot{1}}
\eta_{\dot{2}}\; , \nonumber
\\
\Phi_{1}^{2} + \Phi_{2}^{2} = (\xi^{1} \eta_{\dot{1}}  - \xi^{2}
\eta_{\dot{2}})^{2} - (  \xi^{1} \eta_{\dot{1}} + \xi^{2}
\eta_{\dot{2}}  )^{2} =
  - 4  \; \xi^{1} \xi^{2} \; \eta_{\dot{1}}  \eta_{\dot{2}} \; ,
\nonumber
\\\Phi^{0}  \Phi^{0}  -   \Phi_{1}  \Phi^{j} - \Phi_{2}  \Phi^{2} -
\Phi_{3}  \Phi^{3}= 0 \; ;
\end{eqnarray}

\noindent the complex vector $\Phi_{a}$ is isotropic. Let us
separate real and imaginary parts:
\begin{eqnarray}
\Phi_{0} = A + i B \; , \qquad \Phi_{j} = A_{j} + i B_{j}\; ,
\nonumber
\\
A^{2} -{\bf A}^{2} = B^{2} - {\bf B}^{2}   , \qquad A B -{\bf A}
{\bf B}  = 0\; .
\end{eqnarray}

\noindent So two real 4-vectors $A_{n}$ and $B_{n}$  have the same
length, they are orthogonal to each other, and
 they may be non-isotropic ones. To clarify this, let us detail the structure of them.
The main relationships between spinor and tensors are
\begin{eqnarray}
E  \; ( \Psi \otimes \Psi ) = \gamma^{n}  \; \Phi_{n} + i \;
\sigma^{mn} \; \Phi_{mn} \; , \nonumber
\\
E \; ( \Psi^{*} \otimes \Psi^{*} ) = (\gamma^{n})^{*} \;
\Phi_{n}^{*} - i \; (\sigma^{mn})^{*} \;  \Phi_{mn}^{*} \; .
\label{ad.1}
\end{eqnarray}

\noindent As in spinor basis we have identities
\begin{eqnarray}
(\gamma^{n})^{*} = \gamma^{2} \; \gamma^{n} \; \gamma^{2}\; ,
\qquad (\sigma^{mn})^{*} = - \gamma^{2} \; \sigma^{mn} \;
\gamma^{2} \;, \nonumber
\end{eqnarray}

\noindent relations (\ref{ad.1}) read
\begin{eqnarray}
E  \; ( \Psi \otimes \Psi ) = \gamma^{n}  \; \Phi_{n} + i \;
\sigma^{mn} \; \Phi_{mn} \;,
\nonumber
\\
\gamma^{2} \; E \; ( \Psi^{*} \otimes \Psi^{*} )  \gamma^{2} =
\gamma^{n} \;  \Phi_{n}^{*} + i \;  \sigma^{mn}  \;  \Phi_{mn}^{*}
\; . \label{ad.2}
\end{eqnarray}

\noindent  and further
\begin{eqnarray}
( \Psi \otimes \Psi ) = [\; \gamma^{n}  \; \Phi_{n} + i \;
\sigma^{mn} \; \Phi_{mn} \; ]\; E^{-1}\; ,
 \nonumber
\\
   (   \gamma^{2} \Psi^{*} \otimes   \gamma^{2} \Psi^{*} )  =  [\;  \gamma^{n} \;  \Phi_{n}^{*} +
i \;  \sigma^{mn}  \;  \Phi_{mn}^{*} \; ] \;E^{-1} \; .
\label{ad.3}
\end{eqnarray}

Complex vectors and tensors are given by (see (\ref{9.5.3}))
\begin{eqnarray}
A_{n} +i B_{n} = \Phi _{n}  = {1\over 4} \; \mbox{Sp}  \; [\;  E
\gamma _{n} ( \Psi \otimes \Psi )\;  ] \; , \nonumber
\\
A_{n} - i B_{n} = \Phi _{n}^{*}   =   {1\over 4} \; \mbox{Sp}  \;
[\;
 E \gamma _{n}   (  \gamma^{2} \Psi^{*} \otimes  \gamma^{2} \Psi^{*} ) \; ] \; ,
\nonumber
\\
\Phi _{mn} = -{1 \over 2i} \; \mbox{Sp} \;  [ \; E \sigma _{mn} (
\Psi \otimes \Psi )\; ]\; , \nonumber
\\
\Phi _{mn}^{*}  = - {1 \over 2i} \; \mbox{Sp} \;
 [ \; E \sigma _{mn}  (  \gamma^{2}  \Psi^{*} \otimes  \gamma^{2}  \Psi^{*} ) \; ] \; .
\end{eqnarray}

We may use conventional notation $\gamma^{2} \Psi^{*} = \Psi^{c}$,
then the formulas look shorter
\begin{eqnarray}
A_{n} =
 {1\over 8} \; \mbox{Sp}  \; [\;  E \gamma _{n} ( \Psi \otimes \Psi + \Psi^{c} \otimes   \Psi^{c}  )\;  ]
\; ,
\nonumber
\\
i B_{n} =    {1\over 8} \; \mbox{Sp}  \; [\;
 E \gamma _{n}   (   \Psi^{c} \otimes   \Psi^{c} - \Psi^{c} \otimes   \Psi^{c} ) \; ] \; ,
\nonumber
\\
\Phi _{mn} = -{1 \over 2i} \; \mbox{Sp} \;  [ \; E \sigma _{mn} (
\Psi \otimes \Psi )\; ]\; ,
\nonumber
\\
\Phi _{mn}^{*}  = - {1 \over 2i} \; \mbox{Sp} \;
 [ \; E \sigma _{mn}  (    \Psi^{c} \otimes    \Psi^{c} ) \; ] \; .
\nonumber
\\
\label{shorter}
\end{eqnarray}

Let specify the  complex tensor
\begin{eqnarray}
\Phi^{01} =  {i \over 4}  \;  [ \; (\xi^{1} \xi^{1} - \xi^{2}
\xi^{2}) + ( \eta_{\dot{1}} \eta_{\dot{1}} -  \eta_{\dot{2}}
\eta_{\dot{2}}) \;  ]   \; ,
\nonumber
\\
\Phi^{23} = {1 \over 4} \;   [\;  (\xi^{1} \xi^{1} - \xi^{2}
\xi^{2}) - ( \eta_{\dot{1}} \eta_{\dot{1}} -  \eta_{\dot{2}}
\eta_{\dot{2}}) \;  ]   \; , \nonumber
\\
s^{1} =  \Phi^{01} + i \Phi^{23} = {i \over 2}  \;  (\xi^{1}
\xi^{1} - \xi^{2} \xi^{2}) \; ,
\nonumber
\\
t^{1} =  \Phi^{01} - i \Phi^{23} = {i \over 2 } \; (
\eta_{\dot{1}} \eta_{\dot{1}} -  \eta_{\dot{2}} \eta_{\dot{2}})
\; ; \nonumber
\\
[2mm]
\Phi^{02} = - {1 \over 4}   \; [  \; (\xi^{1} \xi^{1} + \xi^{2}
\xi^{2}) + ( \eta_{\dot{1}} \eta_{\dot{1}} +  \eta_{\dot{2}}
\eta_{\dot{2}})\;   ]\; , \nonumber
\\
\Phi^{31} = - {1 \over 4i}   \; [  \; (\xi^{1} \xi^{1} + \xi^{2}
\xi^{2}) - ( \eta_{\dot{1}} \eta_{\dot{1}} +  \eta_{\dot{2}}
\eta_{\dot{2}})\;   ]\; , \nonumber
\\
s^{2} =  \Phi^{02} + i \Phi^{31} = - {1 \over 2}   \;  (\xi^{1}
\xi^{1} + \xi^{2} \xi^{2}) \; ,
\nonumber
\\
t^{2} =  \Phi^{02} - i \Phi^{31} =
 -{1 \over 2}\; ( \eta_{\dot{1}} \eta_{\dot{1}} +  \eta_{\dot{2}} \eta_{\dot{2}})\; ;
\nonumber
\end{eqnarray}
\begin{eqnarray}
\Phi^{03} = - {i \over 2 }  \;  (  \;  \xi^{1} \xi^{2} +
\eta_{\dot{1}} \eta_{\dot{2}}  ) \; , \qquad
\Phi^{12} = - {1 \over 2 }  \;  (  \;  \xi^{1} \xi^{2} -
\eta_{\dot{1}} \eta_{\dot{2}}  ) \; , \nonumber
\\
s^{3} =  \Phi^{03} + i \Phi^{12} = - i  \;    \xi^{1} \xi^{2} \; ,
\qquad
t^{3} =  \Phi^{03} - i \Phi^{12} =  - i \; \eta_{\dot{1}}
\eta_{\dot{2}}   \; .
\end{eqnarray}

\noindent The vectors ${\bf s}$ and ${\bf t}$ are  isotropic:
\begin{eqnarray}
{\bf s}^{2} = - {1 \over 4}  \;  (\xi^{1} \xi^{1} - \xi^{2}
\xi^{2})^{2} + {1 \over 4}   \;  (\xi^{1} \xi^{1} + \xi^{2}
\xi^{2})^{2} - (\xi^{1} \xi^{2}  )^{2} \equiv 0 \; , \nonumber
\\
{\bf t}^{2} = -{1 \over 4 } \; ( \eta_{\dot{1}} \eta_{\dot{1}} -
\eta_{\dot{2}} \eta_{\dot{2}})^{2} + {1 \over 4}\; (
\eta_{\dot{1}} \eta_{\dot{1}} +  \eta_{\dot{2}}
\eta_{\dot{2}})^{2} - (\eta_{\dot{1}} \eta_{\dot{2}})^{2} \equiv 0
\; ; \nonumber
\end{eqnarray}

\noindent besides
\begin{eqnarray}
{\bf s} \; {\bf t} = -{1 \over 4}\; (\xi^{1} \xi^{1} - \xi^{2}
\xi^{2})  ( \eta_{\dot{1}} \eta_{\dot{1}} -  \eta_{\dot{2}}
\eta_{\dot{2}}) + \nonumber
\\
+ {1 \over 4}\; (\xi^{1} \xi^{1} + \xi^{2} \xi^{2})  (
\eta_{\dot{1}} \eta_{\dot{1}} +  \eta_{\dot{2}} \eta_{\dot{2}}) -
\xi^{1} \xi^{2}\;   \eta_{\dot{1}} \eta_{\dot{2}}= {1 \over 2}(
\xi^{1} \eta_{\dot{2}} -   \xi^{2} \eta_{\dot{1}} )^{2} \; .
\end{eqnarray}

Let us check the sign of the relativistic length for $A_{n}$  (it
equal that for  $B_{n}$):
\begin{eqnarray}
2A_{0} = (\xi^{1}  \eta_{\dot{2}}  - \xi^{2}  \eta_{\dot{1}}) +
     (\xi^{1*}  \eta_{\dot{2}}^{*}  - \xi^{2*}  \eta_{\dot{1}}^{*})\;,
\nonumber
\\
2A_{3} =   - \;  (  \xi^{1} \eta_{\dot{2}} + \xi^{2}
\eta_{\dot{1}}  ) - (  \xi^{1*} \eta_{\dot{2}} ^{*}+ \xi^{2*}
\eta_{\dot{1}} ^{*} ) \; , \nonumber
\\
A_{1} =   (\xi^{1} \eta_{\dot{1}}  - \xi^{2} \eta_{\dot{2}})  +
 (\xi^{1*} \eta_{\dot{1}}^{*}  - \xi^{2*} \eta_{\dot{2}}^{*})  \; ,
 \nonumber
 \\
A_{2} =
   i \;  (  \xi^{1} \eta_{\dot{1}} + \xi^{2} \eta_{\dot{2}}  ) - i
   (  \xi^{1*} \eta_{\dot{1}}^{*} + \xi^{2*} \eta_{\dot{2}} ^{*} ) \; ;
\nonumber
\end{eqnarray}

\noindent allowing for identities
\begin{eqnarray}
4( A^{2}_{0} - A_{3}^{2})=  - 4 \xi^{1} \xi^{2} \;  \eta_{\dot{1}}    \eta_{\dot{2}} - 4
\xi^{1*}  \xi^{2*} \;   \eta_{\dot{1}}^{*} \eta_{\dot{2}}^{*} - 4
\xi^{1} \xi^{1*} \; \eta_{\dot{2}}  \eta_{\dot{2}}^{*} - 4
\xi^{2} \xi^{2*} \; \eta_{\dot{1}}  \eta_{\dot{1}}^{*} \; ,
\nonumber
\\
4 ( A_{1}^{2} + A_{2}^{2} )=  - 4 \xi^{1} \xi^{2} \;  \eta_{\dot{1}}    \eta_{\dot{2}} - 4
\xi^{1*}  \xi^{2*} \;   \eta_{\dot{1}}^{*} \eta_{\dot{2}}^{*} + 4
\xi^{1} \xi^{1*} \; \eta_{\dot{1}}  \eta_{\dot{1}}^{*} + 4
\xi^{2} \xi^{2*} \; \eta_{\dot{2}}  \eta_{\dot{2}}^{*} \; ,
 \nonumber
 \end{eqnarray}

 \noindent we arrive at
 \begin{eqnarray}
 A^{2}_{0} - {\bf A}^{2} =  - (\xi^{1} \xi^{1*}  + \xi^{2} \xi^{2*}) \; \; ( \eta_{\dot{1}}
\eta_{\dot{1}}^{*} +  \eta_{\dot{2}}  \eta_{\dot{2}}^{*} ) < 0 \;.
\end{eqnarray}

\noindent Therefore, this 4-vector is space-like, and it cannot
correspond to  a time-like Stoke 4-vector.

\section{Spinor representation for a time-like vectors, \\ on possible Jones spinor for a partly polarized light }

Now let us examine else one possibility
\begin{eqnarray}
\Psi \otimes (-i \Psi^{c} )  = \left | \begin{array}{c}
 \xi^{1} \\  \xi^{2} \\
\eta_{\dot{1}} \\ \eta_{\dot{2}}
\end{array} \right |  \otimes \left | \begin{array}{c}
+  \eta_{\dot{2}}^{*} \\ - \eta_{\dot{1}}^{*} \\ -  \xi^{2*} \\ +
\xi^{1*}
\end{array} \right |= \qquad \qquad
\nonumber
\\
= \left | \begin{array}{cccc}
+  \xi^{1}\eta_{\dot{2}}^{*}  & -\xi^{1}  \eta_{\dot{1}}^{*} & -  \xi^{1}\xi^{2*} &  + \xi^{1} \xi^{1*} \\
+  \xi^{2} \eta_{\dot{2}}^{*}  & - \xi^{2} \eta_{\dot{1}}^{*} & -  \xi^{2} \xi^{2*} &  + \xi^{2} \xi^{1*} \\
+ \eta_{\dot{1}} \eta_{\dot{2}}^{*}  & - \eta_{\dot{1}} \eta_{\dot{1}}^{*} & -  \eta_{\dot{1}} \xi^{2*} &  + \eta_{\dot{1}}\xi^{1*} \\
+  \eta_{\dot{2}}\eta_{\dot{2}}^{*}  & - \eta_{\dot{2}}\eta_{\dot{1}}^{*} & -\eta_{\dot{2}}  \xi^{2*} &  + \eta_{\dot{2}}\xi^{1*} \\
\end{array} \right |=
\left | \begin{array}{cccc}
\xi^{11} &  \xi^{12}  & \Delta^{1}_{\;\;\dot{1}}    &   \Delta ^{1}_{\;\;\dot{2}} \\
\xi^{21} &  \xi^{22}  & \Delta ^{2}_{\;\;\dot{1}}   &   \Delta ^{2}_{\;\;\dot{2}} \\
H_{\dot{1}} ^{\;\;1}  &  H_{\dot{1}}^{\;\;2}  & \eta _{\dot{1} \dot{1}}  & \eta_{\dot{1} \dot{2}} \\
H_{\dot{2}}^{\;\;1}  &  H_{\dot{2}}^{\;\;2}  &
\eta_{\dot{2}\dot{1}}  & \eta_{\dot{2}\dot{2}}
 \end{array} \right | \; .
\end{eqnarray}

\noindent Corresponding 4-vector is determined  by
\begin{eqnarray}
\Phi_{0} =  {1 \over 2} \; [ \; (H_{\dot{2}}^{\;\; 1} -
H_{\dot{1}}^{\;\; 2} ) - ( \Delta^{2}_{\;\;\dot{1}}  -
\Delta^{1}_{\;\;\dot{2}} ) \; ] = {1 \over 2} \; [\; (
\eta_{\dot{2}}  \eta_{\dot{2}}^{*}  +  \eta_{\dot{1}}
\eta_{\dot{1}}^{*}) + (  \xi^{2} \xi^{2*} + \xi^{1} \xi^{1*} )\; ]
> 0 \; , \nonumber
\\
\Phi_{3} =  -{ 1 \over 2}\; [ \;  ( H_{\dot{2}}^{\;\;1} +
H_{\dot{1}}^{\;\;2} ) +
 (\Delta^{2}_{\;\;\dot{1}} + \Delta^{1}_{\;\;\dot{2}}  ) \; ] =
 -  {1 \over 2} \; [\; ( \eta_{\dot{2}}\eta_{\dot{2}}^{*} - \eta_{\dot{1}} \eta_{\dot{1}}^{*} ) +
 (-  \xi^{2} \xi^{2*}+   \xi^{1} \xi^{1*}  )\; ] \; ,
\nonumber
\\
\Phi_{1} =   {1 \over 2}\; [ \; (H_{\dot{1}}^{\;\;1} -
H_{\dot{2}}^{\;\;2}) + (\Delta^{1}_{\;\;\dot{1}} -
\Delta^{2}_{\;\;\dot{2}} )\; ] = {1 \over 2} \;  [\;
 (\eta_{\dot{1}} \eta_{\dot{2}}^{*} +\eta_{\dot{2}}\eta_{\dot{1}}^{*} ) -
 (  \xi^{1}\xi^{2*} + \xi^{2} \xi^{1*}  ) \; ]  \; ,
\nonumber
\\
\Phi_{2} =
 { i \over 2}\; [  \; ( H_{\dot{1}}^{\;\;1} + H_{\dot{2}}^{\;\;2} ) +
 (\Delta^{1}_{\;\;\dot{1}} + \Delta^{2}_{\;\;\dot{2}}  ) \; ] =
   {i   \over 2}  \;  [ ( \eta_{\dot{1}} \eta_{\dot{2}}^{*}  - \eta_{\dot{2}}\eta_{\dot{1}}^{*}    )  +
    (-  \xi^{1}\xi^{2*} +  \xi^{2} \xi^{1*} )\; ] \; .
\nonumber
\end{eqnarray}

\noindent Allowing for
\begin{eqnarray}
4 (\Phi_{0}^{2} - \Phi_{3}^{2} )  =  4 \eta_{\dot{1}}\eta_{\dot{1}}^{*}  \;
\eta_{\dot{2}}\eta_{\dot{2}}^{*}  + 4  \xi^{1} \xi^{1*} \;
\xi^{2} \xi^{2*} + 4 \eta_{\dot{1}}  \eta_{\dot{1}}^{*}\; \xi^{1}
\xi^{1*} + 4 \eta_{\dot{2}}  \eta_{\dot{2}}^{*}\; \xi^{2} \xi^{2*}
\; ,
\nonumber
\\
4 (\Phi_{1}^{2} + \Phi_{2}^{2} )  = 4 \eta_{\dot{1}} \eta_{\dot{1}}^{*} \;  \eta_{\dot{2}}
\eta_{\dot{2}}^{*} \;  + 4 \xi^{1} \xi^{1*} \;   \xi^{2} \xi^{2*}
 -4 \eta_{\dot{1}} \eta_{\dot{2}}^{*} \; \xi^{2} \xi^{1*} -4 \eta_{\dot{2}}\eta_{\dot{1}}^{*} \xi^{1}\xi^{2*} \; ,
\nonumber
\end{eqnarray}

\noindent we get
\begin{eqnarray}
\Phi^{a} \Phi_{a}= \Phi_{0}^{2} - \Phi_{1}^{2} -  \Phi_{2}^{2} -
\Phi_{1}^{3} = \nonumber
\\
=
 \eta_{\dot{1}}  \eta_{\dot{1}}^{*}\; \xi^{1} \xi^{1*} +  \eta_{\dot{2}}  \eta_{\dot{2}}^{*}\; \xi^{2} \xi^{2*}
+ \eta_{\dot{1}} \eta_{\dot{2}}^{*} \; \xi^{2} \xi^{1*}
+\eta_{\dot{2}}\eta_{\dot{1}}^{*} \xi^{1}\xi^{2*}  \; .
\end{eqnarray}

\noindent
Let us demonstrate that this vector is time-like. With the
notation
\begin{eqnarray}
\xi = \left | \begin{array}{c}
N_{1} e^{in_{1}} \\
N_{2} e^{in_{2}} \end{array} \right |\; , \qquad \eta = \left |
\begin{array}{c}
M_{1} e^{im_{1}} \\
M_{2} e^{im_{2}} \end{array} \right |\; , \label{Jonesspinors}
\end{eqnarray}

\noindent we get
\begin{eqnarray}
\Phi^{a} \Phi_{a} = N_{1}^{2} M_{1}^{2} +  N_{2}^{2} M_{2}^{2} + 2
N_{1}M_{1} \; N_{2}M_{2} \; \cos \;  [( n_{1}-n_{2}) - (m_{1}
-m_{2} )] \; ; \nonumber
\end{eqnarray}

\noindent therefore
\begin{eqnarray}
(N_{1}  M_{1}  -  N_{2}  M_{2})^{2}  < \Phi_{0}^{2} - \Phi_{1}^{2}
-
 \Phi_{2}^{2} - \Phi_{1}^{3}  < (N_{1}  M_{1}  +  N_{2}  M_{2})^{2} \; .
\end{eqnarray}

This means that we have ground to consider   4-vector  $\Phi_{a}$
as Stokes 4-vector $S_{a}$:
\begin{eqnarray}
(N_{1}  M_{1}  -  N_{2}  M_{2})^{2}  < S_{0}^{2} - {\bf S}^{2}   <
(N_{1}  M_{1}  +  N_{2}  M_{2})^{2} \; ,
\end{eqnarray}

\noindent and two 2-spinors (\ref{Jonesspinors})  as making up a
Jones bi-spinor corresponding a partly polarized light.

It remains to find explicit form  for corresponding (real) Stokes
4-tensor $S_{ab}$:
\begin{eqnarray}
\Phi^{01} = {i \over 4} \;   [ \; ( \xi^{1}\eta_{\dot{2}}^{*}  +
\xi^{2} \eta_{\dot{1}}^{*}) - (   \eta_{\dot{1}} \xi^{2*}    +
\eta_{\dot{2}}\xi^{1*} ) \;  ] \; , \nonumber
\\
\Phi^{23} = {1 \over 4}\; [ \; ( \xi^{1}\eta_{\dot{2}}^{*}  +
\xi^{2} \eta_{\dot{1}}^{*}) + (   \eta_{\dot{1}} \xi^{2*}    +
\eta_{\dot{2}}\xi^{1*} ) \;  ]\; , \nonumber
\\
\Phi^{02} = - {1 \over 4} \;   [  \; (\xi^{1}\eta_{\dot{2}}^{*} -
\xi^{2} \eta_{\dot{1}}^{*}) +
 ( -  \eta_{\dot{1}} \xi^{2*}  +  \eta_{\dot{2}}\xi^{1*} )  \; ] \; ,
\nonumber
\\
\Phi^{31} =
 {i \over 4} \;   [  \; (\xi^{1}\eta_{\dot{2}}^{*} - \xi^{2} \eta_{\dot{1}}^{*}) -
 ( -  \eta_{\dot{1}} \xi^{2*}   +  \eta_{\dot{2}}\xi^{1*} )  \; ] \; ,
\nonumber
\end{eqnarray}
\begin{eqnarray}
\Phi^{03} = - {i \over 4}  \;  [ \;  (\xi^{2} \eta_{\dot{2}}^{*}
-\xi^{1}  \eta_{\dot{1}}^{*} ) +
 ( -\eta_{\dot{2}}  \xi^{2*}  +  \eta_{\dot{1}}\xi^{1*} )  ]\; ,
\nonumber
\\
\Phi^{12} = - {1 \over 4}  \;  [ \;  (\xi^{2} \eta_{\dot{2}}^{*}
-\xi^{1}  \eta_{\dot{1}}^{*}) - ( -\eta_{\dot{2}}  \xi^{2*}  +
\eta_{\dot{1}}\xi^{1*} )  ] \; ;
\nonumber
\\
s^{1} =a^{1} + i b^{1} = {i \over 2} \;    (
\xi^{1}\eta_{\dot{2}}^{*}  + \xi^{2} \eta_{\dot{1}}^{*}) \; ,
\nonumber
\\
s^{2} =a^{2} + i b^{2} = - {1 \over 2} \;
(\xi^{1}\eta_{\dot{2}}^{*} - \xi^{2} \eta_{\dot{1}}^{*})  \; ,
\nonumber
\\
s^{3} =a^{3} + i b^{3} = - {i \over 2}  \;  (\xi^{2}
\eta_{\dot{2}}^{*}  -\xi^{1}  \eta_{\dot{1}}^{*} ) \; ;
\end{eqnarray}

\noindent besides this complex 3-vector is not isotropic:
$$
{\bf s}^{2} = -{1 \over 4}\; ( \xi^{1} \eta_{1}^{*} - \xi^{2}
\eta_{1}^{*} )^{2} \neq 0 \; .
$$

Last remark should be added. Results of the present consideration  can be of use
not only in  polarization optics, but also they may be of interest
to describe Maxwell theory in spinor approach, when instead
variables $A_{n}, F_{mn}$ one introduces one  fundamental
electromagnetic bi-spinors $\Psi = (\xi , \eta)$. Also, they could
have meaning in the context of explicit constructing  models for
space-time with spinor structure.

\section*{Acknowledgements}

Author is grateful  to  participants of seminar of Laboratory of
theoretical physics,
 National Academy of Sciences of Belarus for discussion.
 Authors are grateful to an anonymous reviewer for comments and advice stimulated improving
 the paper.

This  work was  supported  by Fund for Basic Research of Belarus
 F09K-123.

\end{document}